\documentclass[5p]{elsarticle}   %use either "review" (for single column) or "5p" for double column format

\usepackage{lineno,hyperref,graphicx,amsmath,float,color}
\modulolinenumbers[5]

\journal{Optics Communications}

\biboptions{numbers,sort&compress} %this changes a citation from [1,2,3] to [1-3] for example.

\begin{document}

\begin{frontmatter}

\title{Feasibility of Single-Photon Cross-Phase Modulation using Metastable Xenon in a High Finesse Cavity}

%% Group authors per affiliation:
\author{B. T. Kirby*, G. T. Hickman, T. B. Pittman and J. D. Franson}
\address{Physics Department, University of Maryland Baltimore County, Baltimore, MD 21250}
\cortext[cor1]{Corresponding author. Tel.:+1-410-455-2526. \\ \textit{E-mail address:} ki6@umbc.edu (B. T. Kirby)}

%% or include affiliations in footnotes:
%\author[mymainaddress,mysecondaryaddress]{Elsevier Inc}
%\ead[url]{www.elsevier.com}

%\author[mysecondaryaddress]{Global Customer Service\corref{mycorrespondingauthor}}
%\cortext[mycorrespondingauthor]{Corresponding author}
%\ead{support@elsevier.com}

%\address[mymainaddress]{1600 John F Kennedy Boulevard, Philadelphia}
%\address[mysecondaryaddress]{360 Park Avenue South, New York}

\begin{abstract}
Cross-phase modulation at the single-photon level has a wide variety of fundamental applications in quantum optics including the generation of macroscopic entangled states. Here we describe a practical method for producing a weak cross-phase modulation at the single-photon level using metastable xenon in a high finesse cavity. We estimate the achievable phase shift and give a brief update on the experimental progress towards its realization. A single-photon cross-phase modulation of approximately 20 milliradians is predicted by both a straightforward perturbation theory calculation and a numerical matrix diagonalization method.
\end{abstract}


\end{frontmatter}

\section{Introduction}

A wide variety of proposed experiments in quantum optics make use of cross-phase modulation at the single-photon level.
It is particularly useful in the creation of Schrodinger cat states and entangled coherent states which have applications in quantum computing \cite{Munro2005137,Nemoto2004250502,Jeong2002042305}, teleportation \cite{Enk2001022313,Wang2001022302,Jeong2001052308}, metrology \cite{Joo2011083601}, cryptography \cite{Simon2014012315}, and in nonlocal interferometry \cite{Kirby2013053822,Kirby2013ARXIV}.
Nonclassical effects involving entangled coherent states are also useful for probing the boundary between classical and quantum behavior.  
Coherent states are the closest approximation to a classical state of light, making superpositions of sufficiently orthogonal coherent states a truly macroscopic quantum phenomenon.

Experiments to create cross-phase modulation at the single-photon level have been performed with many different technologies and nonlinear media.
Single atoms in micro-cavities have been used \cite{Kimble19954710}, as well as atomic vapor in a hollow core fiber \cite{Gaeta2013138}, transmons at microwave wavelengths \cite{Hoi2013053601}, and a variety of systems using electromagnetically induced transparency \cite{Molella2008273,Chen2011193006,Chen2011041804}.
Other efforts have used quantum dots in a cavity \cite{Fushman2008769} and strongly interacting Rydberg atoms \cite{firstenberg201371}.
Large per photon phase shifts have been measured in many of these systems, but they require relatively complicated experimental setups, prompting a search for a simpler and more reliable source of low power cross-phase modulation.

Here we discuss the feasibility of a new cavity approach for single-photon level cross-phase modulation that uses metastable xenon atoms as the nonlinear medium. 
Meta-stable xenon is expected to be superior to alkali vapors such as rubidium and cesium since it is inert and does not adhere to optical surfaces \cite{Pittman2013053804}. 
Xenon also has a long metastable lifetime and relatively large dipole matrix elements with a convenient set of ladder transitions in the near infrared.
The two level spacings are relatively close in wavelength, allowing approximately Doppler-free experiments with counterpropagating beams.

The use of a high finesse cavity should avoid the limitations in using freely-propagating beams that have been pointed out by Shapiro and others based on a multi-mode analysis \cite{Shapiro200716, Gea-Banacloche2010043823,Fan2013053601}.  
These difficulties do not occur as long as only a single cavity resonant frequency exists within the bandwidth of the medium.

We estimate that a single photon in the setup described here will be able to produce a nonlinear phase shift of approximately 20 milliradians, as described in section \ref{sec:theory}.
Two different methods for calculating the magnitude of the expected cross-phase modulation are found to be in good agreement.
One of these consists of a straightforward analytical calculation based on perturbation theory.
Those results are then verified using a numerical matrix diagonalization method which is more appropriate for large numbers of atoms and small detunings.
In section \ref{sec:experiment} we briefly describe the progress of an ongoing experimental effort towards the realization of the metastable xenon approach.
Finally in section \ref{sec:summary} we provide a conclusion and summary of results.

\section{Theoretical Model}
\label{sec:theory}

\subsection{Three-Level System}
\label{subsec:theorysetup}

\begin{figure}[tbp]
\begin{center}
\includegraphics[height=5cm,width=7cm,angle=0]{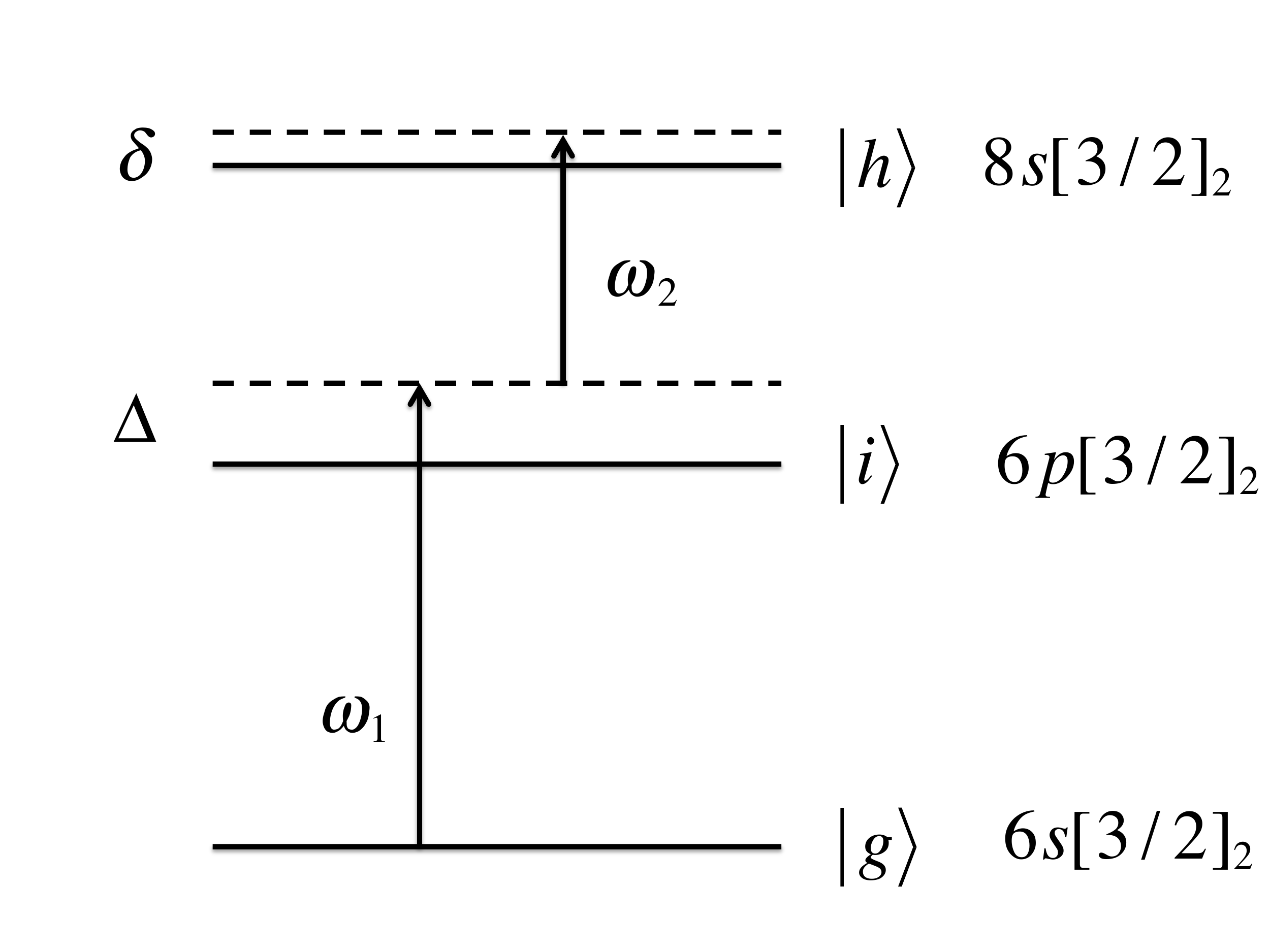}
\end{center}
\caption{Three-level system in metastable xenon used to generate cross-phase modulation.  The detunings of the control and signal ($\omega_1$ and $\omega_2$) are given by $\Delta$ and $\delta$ respectively.  The atomic levels are represented by $|g\rangle$, $|i\rangle$, and $|h\rangle$, where $|g\rangle$ is the metastable state with an intrinsic lifetime of approximately 43 seconds \cite{Walhout29942843}.  The transitions of interest correspond to wavelengths of 823 and 853 nm for the control and signal respectively \cite{Allen1969842,Das2005285}.}
\label{fig:detunings}
\end{figure}

The xenon transitions of interest form a three-level ladder system as pictured in Fig. \ref{fig:detunings}, where $\omega_1$ and $\omega_2$ represent the control and signal photon frequencies, respectively.
Two-photon absorption can be minimized by detuning the signal and control photons from atomic resonance, so that the net effect is a conditional nonlinear phase shift.
For applications involving the generation of phase-entangled coherent states \cite{Kirby2013053822,Kirby2013ARXIV}, the control at $\omega_{1}$ would be a single photon while the signal at $\omega_2$ would be a weak coherent state.

Three-level systems of this kind have previously been analyzed using a density matrix approach \cite{You2008053803}.  
Our goal here is to use a straightforward perturbation calculation to obtain an approximate estimate of the cross-phase modulation in metastable xenon, which we can use to demonstrate the feasibility of the approach.

We define a set of basis states for describing the interaction in Fig. \ref{fig:detunings} as
 \begin{equation}
 \begin{split}
|1\rangle = & |0\rangle \otimes |h\rangle \\
|2\rangle = & \hat{a}^{\dagger}_{\omega_{1}}|0\rangle \otimes |i\rangle \\
|3\rangle = & \hat{a}^{\dagger}_{\omega_{2}}|0\rangle \otimes |i\rangle \\
|4\rangle = & \hat{a}^{\dagger}_{\omega_{1}}\hat{a}^{\dagger}_{\omega_{2}}|0\rangle \otimes |g\rangle. \\
  \end{split}
    \label{eq:basis}
\end{equation}
Here $\hat{a}^{\dagger}_{\omega_{i}}$ is the usual creation operator for angular frequency $\omega_i$ and $|0\rangle$ is the vacuum state of the field.
In this basis the interaction Hamiltonian $\hat{V}$ can be defined in the usual way as \cite{Scully1997}:
 \begin{equation}
 \begin{split}
\hat{V}&= m_{1}^{*}\hat{\sigma}^{\dagger}_{gi}\hat{a}_{\omega_{1}}+m_{2}^{*}\hat{\sigma}^{\dagger}_{ih}\hat{a}_{\omega_{2}}+m_{1}\hat{\sigma}_{gi}\hat{a}^{\dagger}_{\omega_{1}}+m_{2}\hat{\sigma}_{ih}\hat{a}^{\dagger}_{\omega_{2}}\\
&+m_{3}^{*}\hat{\sigma}_{gi}^{\dagger}\hat{a}_{\omega_{2}}+m_{3}\hat{\sigma}_{gi}\hat{a}^{\dagger}_{\omega_{2}},
  \end{split}
    \label{eq:Vdef}
\end{equation}
where $\hat{\sigma}_{gi}$ takes the atom from $|i\rangle$ to $|g\rangle$, $\hat{\sigma}_{ih}$ takes the atom from $|h\rangle$ to $|i\rangle$ and the $m$ terms are the transition matrix elements.
In general the matrix elements are given by $m=\langle - \vec{\mu} \cdot \vec{E} \rangle$ where $\mu$ is the dipole moment of the transition and $\vec{E}$ is the electric field, and the brackets indicate an average over orientations.

The basis states of Eq. (\ref{eq:basis}) and the interaction Hamiltonian of Eq. (\ref{eq:Vdef}) describe a system that can undergo several kinds of transitions.
The system may initially transition from state $|4\rangle$ to either state $|3\rangle$ or state $|2\rangle$ by the absorption of the control ($\omega_1$) or signal ($\omega_2$) photons respectively.
A second photon may then be absorbed to take the system from states $|2\rangle$ or $|3\rangle$ to state $|1\rangle$ \cite{You2008053803,You20121627}.
Using Eqs. (\ref{eq:basis}) and (\ref{eq:Vdef}) the total Hamiltonian $\hat{H}$ of the system can be written as
   \begin{equation}
   \setlength{\arraycolsep}{1pt}
   \hat{H}=
 \begin{pmatrix}
 \hbar(\omega_{hi}+\omega_{ig}) & 0 & m_{2}^{*} & 0\\
 0 &  \hbar(\omega_{1}+\omega_{ig}) &  0 & m_{3}^{*} \\
  m_{2} & 0 & \hbar(\omega_{2}+\omega_{ig}) & m_{1}^{*} \\
  0 & m_{3} & m_{1} &  \hbar(\omega_{1}+\omega_{2})
 \end{pmatrix}
 \label{eq:totalh}
 \end{equation}

The finite lifetimes of the excited levels are not taken into account here in order to keep the presentation as transparent as possible.
Inclusion of the lifetimes reduces the cross-phase modulation by an amount that is not significant for detunings much larger than the line width, as is expected to be the case in the planned experiments.
The intrinsic lifetime of the metastable $6s[3/2]_2$ state (approximately 43 seconds \cite{Walhout29942843}) has no significant effect on the results.

\subsection{Perturbation Theory}
\label{subsec:pert}

A straightforward perturbation theory approach can be used to estimate the cross-phase shift for sufficiently large detunings.
In that limit, the nonlinear phase shift can be calculated for a single atom and then summed over the contributions from all of the atoms.
This approach is valid as long as the depopulation of the initial state is sufficiently small, as will be verified below using a numerical diagonalization technique. 

The level spacings and detunings are chosen in such a way that the control photon effectively interacts only with levels $|g\rangle$ and $|i\rangle$ while the signal photon only interacts with levels $|i\rangle$ and $|h \rangle$.
To fourth order in perturbation theory, each photon is absorbed and re-emitted once, returning the atom back to the ground state.  
The assumption that only this 4th-order term is necessary to predict the phase shift is confirmed by the numerical approach of section \ref{subsec:monte}, which takes all orders into account.

The fourth order term of interest gives a change $E^{(4)}$ in the energy of the system given by  \cite{You2008053803}:
 \begin{equation}
E^{(4)}=\frac{|m_{1}|^{2}|m_{2}|^{2}}{\hbar^{3}\Delta^2\delta}.
    \label{eq:energyshift}
\end{equation}
The matrix elements are a function of position within the cavity, which is assumed to contain a uniform density $\rho$ of metastable xenon atoms.
The total phase shift from a single atom is determined by the fact that the time dependence of the state is proportional to $exp[-iE^{(4)}t/\hbar]$, which gives a total phase shift of
 \begin{equation}
\phi = \rho \int  \frac{E^{(4)}t}{\hbar} dV.
 \label{eq:phidensity}
 \end{equation}
 Here the integral is over the cavity volume and t is the interaction time inside the cavity.

The integral of Eq. (\ref{eq:phidensity}) can be simplified by making two approximations regarding the electric field within the cavity.
First we replace the sinusoidally varying field with a suitable average, since the field oscillates on a length scale much smaller than the size of the cavity.
This average is found by normalizing the energy of the electric field in the cavity to that of a single photon of the proper wavelength.
Secondly we model the cavity mode field distribution as a constant electric field over a cylinder with a diameter equal to the gaussian beam diameter and length equal to that of the cavity.
These approximations are made only to simplify the presentation.
Numerical integration of Eq. (\ref{eq:phidensity}) using the exact field distribution calculated from the geometry of the cavity \cite{Boyd19611538} is in good agreement and will be discussed in section \ref{subsec:compare}.

The average total phase shift can now be written as
 \begin{equation}
\phi \approx \rho V_{cyl} \frac{|\bar{m}_{1}|^{2}|\bar{m}_{2}|^{2}}{\hbar^{4}\Delta^2\delta}t,
    \label{eq:energyshiftaverage}
\end{equation}
where $V_{cyl}$ is the volume of the cylinder we have used to model the cavity mode field distribution, and $\bar{m}_{i}$ represents the matrix element defined in section \ref{subsec:theorysetup} where the electric field has been averaged over the cavity field distribution.
In addition, the number of atoms involved in the interaction is effectively reduced to $1/3$ of the number present in the ensemble due to averaging of the electric field and dipole moment orientations.

The range of validity of this perturbation calculation is limited to situations in which the total probability that an atom is in an excited state is much less than 1, since perturbation theory does not include the depletion of the initial state.
In practice the population of the first excited state is the limiting factor.
The total number of atoms $N_{1}$ in the first excited state is given approximately by \cite{You20121627}.
 \begin{equation}
N_{1} \approx \rho V_{cyl} \frac{|\bar{m}_{1}|^{2}}{\hbar^{2}\Delta^2}.
    \label{eq:firstlevel}
\end{equation}
Equation (\ref{eq:energyshiftaverage}) is only valid if $N_1$ is significantly less than 1.

\subsection{Matrix Diagonalization}
\label{subsec:monte}

In addition to the simplified perturbation theory calculation of the previous section a numerical matrix diagonalization method was used to verify the results and determine its range of validity.
In this approach the Hamiltonian of Eq. (\ref{eq:totalh}) was diagonalized numerically to determine its eigenvalues.  
The nonlinear energy shift due to cross-phase modulation was then found by subtracting out the linear terms that appear when either the signal or the probe is present by itself.  

Though the mathematics in this approach are not as transparent as for the perturbation theory method of section \ref{subsec:pert}, it has two significant advantages.
The first is that this technique includes all orders of the interaction, unlike the perturbation theory calculation which only took into account fourth-order terms.
The second is that the combined effects of an ensemble of atoms can be properly treated using this approach rather than simply summing over the phase shifts calculated for single atoms.  
This produces accurate results at arbitrarily small detunings, in contrast to the results of section \ref{subsec:pert} which are only valid when the first level population expressed in Eq. (\ref{eq:firstlevel}) is much less than one.

The ensemble of atoms can be taken into account by considering the form of the initial and virtual states in the presence of $N$ identical atoms.
For simplicity, we assume that all of the atoms are coupled to the field in the same way, as in the cylindrical field model described above.
In that case all of the atoms are excited with the same probability amplitude, so that the only possible form of the initial and intermediate states is given by
 \begin{equation}
 \begin{split}
|1'\rangle = & |0\rangle \otimes \frac{1}{\sqrt{N}} \sum_{k=1}^{N}\prod_{j\neq k}^{N}|g_{j}\rangle|h_{k}\rangle \\
|2'\rangle = & \hat{a}^{\dagger}_{\omega_{1}}|0\rangle \otimes \frac{1}{\sqrt{N}} \sum_{k=1}^{N}\prod_{j\neq k}^{N}|g_{j}\rangle|i_{k}\rangle \\
|3'\rangle = & \hat{a}^{\dagger}_{\omega_{2}}|0\rangle \otimes \frac{1}{\sqrt{N}} \sum_{k=1}^{N}\prod_{j\neq k}^{N}|g_{j}\rangle|i_{k}\rangle \\
|4'\rangle = & \hat{a}^{\dagger}_{\omega_{1}}\hat{a}^{\dagger}_{\omega_{2}}|0\rangle \otimes \prod_{j=1}^{N}|g_{j}\rangle. \\
  \end{split}
    \label{eq:basisprime}
\end{equation}
Here the indices $j$ and $k$ label the atoms and $|g_{i}\rangle$, $|i_{i}\rangle$, and $|h_{i}\rangle$ are the three atomic states of atom $i$.

Making the same constant field mode approximation as discussed in section \ref{subsec:pert} we can evaluate the new transition matrix elements using this basis:
 \begin{equation}
 \begin{split}
m_{1}' &= \langle 4' | \hat{V} | 3' \rangle =  \sqrt{N}m_1\\
m_{2}' &= \langle 3' | \hat{V} | 1' \rangle = m_2\\
m_{3}' &= \langle 4' | \hat{V} | 2' \rangle = \sqrt{N}m_3.\\
  \end{split}
    \label{eq:mprimeelements}
\end{equation}
It can be seen that a factor of $\sqrt{N}$ automatically appears in some of the matrix elements.
Note that all $N$ atoms in the ensemble are available to make transitions from $| g \rangle$ to $| i \rangle$, so the matrix elements for these transitions are multiplied by $\sqrt{N}$.
At any given time, however, no more than one atom can occupy state $| i \rangle$, since the excitation is driven by a single photon. As a result the matrix element for the $| i \rangle$ to $| h \rangle$ transition is not affected by an increase in the number of atoms.

This approach neglects inhomogeneous broadening, such as the Doppler shift, which would result in different excited-state amplitudes for different atoms.
But it does provide a qualitative description of the effects of using a large ensemble of atoms.
Doppler shifts become important in the limit of large detunings.

\subsection{Comparison of Calculated Results}
\label{subsec:compare}

The cavity used in these calculations was assumed to consist of a set of two 2.5 cm spherical mirrors placed 2.5 mm apart with a finesse of approximately 60,000.  
The beam waist in this case is approximately 38 $\mu$m and the interaction time between the intracavity medium and a single intracavity photon is roughly 330 ns.
We assumed a metastable xenon density in the cavity of $10^{10}$/cm$^{3}$, though densities of up to $10^{13}$/cm$^{3}$ are possible \cite{Uhm2008211501}.

The relevant xenon transitions are the 823 nm and 853 nm lines for the control and signal respectively \cite{Allen1969842,Das2005285}.
The dipole moments of these transitions can be calculated from their Einstein A coefficients \cite{Hilborn1982982}.  
They are found to be $2.3\times10^{-29}$ C$\cdot$m and $6.4\times10^{-30}$ C$\cdot$m for the first and second transitions respectively.

\begin{figure}[btp]
\begin{center}
\includegraphics[height=6cm,width=9cm,angle=0]{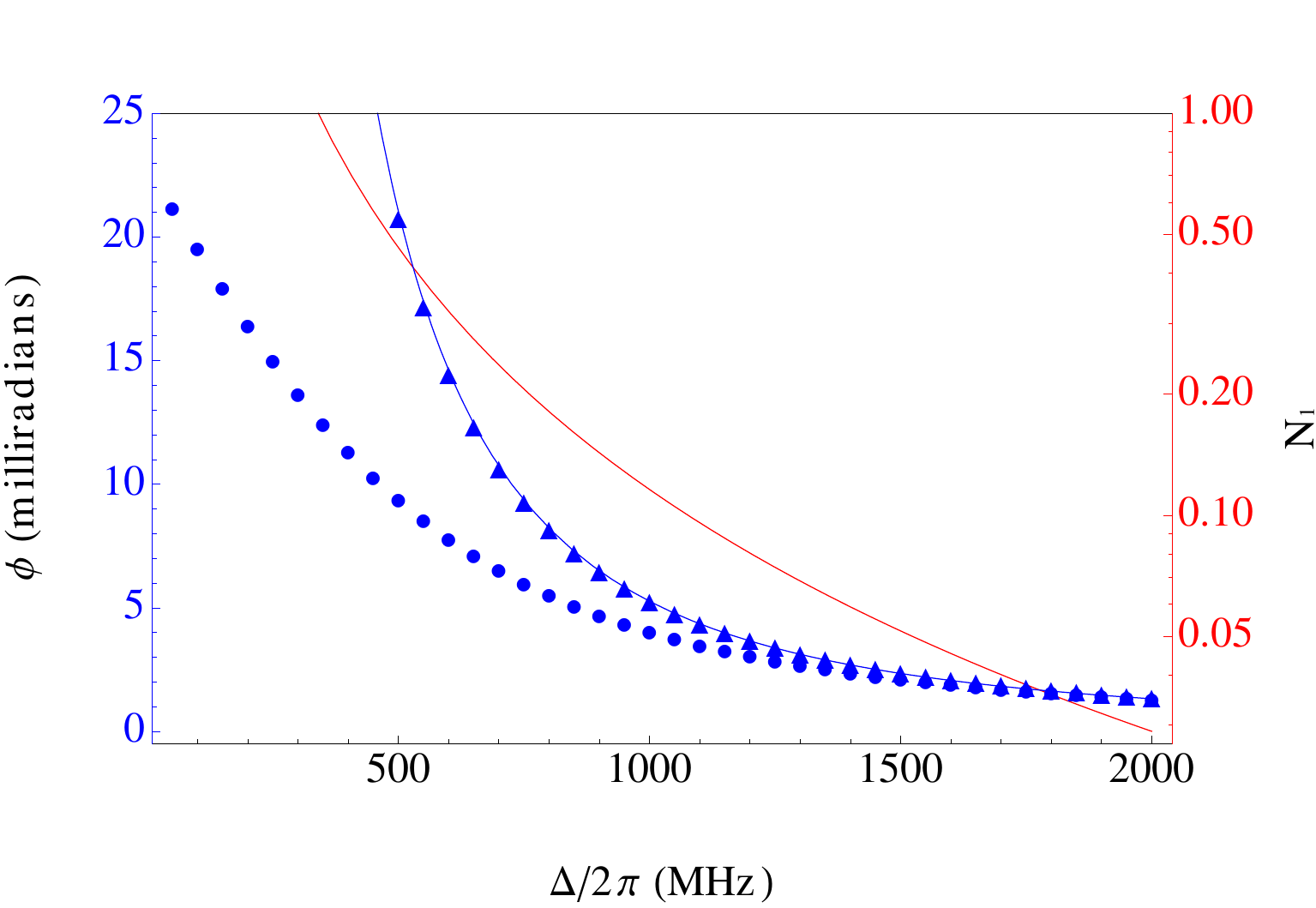}
\end{center}
\caption{Cross-phase modulation values (blue) and first excited state populations (red) as a function of the detuning $\Delta$ from the first transition (the detuning $\delta$ from the upper atomic level is fixed at $\delta/2\pi$ = 10 MHz).  The solid blue line shows the analytical perturbation theory results of Eq. (\ref{eq:energyshiftaverage}), and blue triangles show the results of integrating Eq. \ref{eq:phidensity} numerically over an exact field distribution \cite{Boyd19611538}.
The blue dots show the numerical results of the matrix diagonalization method of section \ref{subsec:monte}. }
\label{fig:phiandN}
\end{figure}

Fig. \ref{fig:phiandN} plots the cross-phase modulation predicted by each approach as a function of the detuning $\Delta$ from the first transition with the upper level detuning $\delta$ held constant at $\delta/2\pi$=10 MHz.
The solid blue line represents the results of the perturbative approach of Eq. (\ref{eq:energyshiftaverage}), in which the electric field mode has been approximated as constant within the cavity mode volume. 
The solid blue triangles represent the results of numerically integrating equation \ref{eq:phidensity} over the exact field distribution (a Gaussian) calculated from the geometry of the cavity \cite{Boyd19611538}.  
The close agreement of these two approaches justifies the use of the constant field mode approximation.
Results from the matrix diagonalization approach of section \ref{subsec:monte} are represented as blue dots.

The red line in Fig. \ref{fig:phiandN} corresponds to the number $N_{1}$ of atoms in the first excited state as a function of the intermediate state detuning $\Delta$, as calculated from Eq. (\ref{eq:firstlevel}).
The results of the two approaches diverge for large $N_{1}$ as expected.
From the results of the matrix diagonalization approach we see that the largest single-photon cross-phase modulation possible for this particular cavity setup is approximately 20 milliradians.
Phase shifts of this magnitude should be sufficient for a number of applications using large-amplitude coherent states and weak nonlinearities, such as the generation of Schrodinger cat states \cite{Jeong2004061801, Gerry19994095}, nonlocal interferometry \cite{Kirby2013053822,Kirby2013ARXIV}, quantum computing \cite{Munro2005137,Nemoto2004250502}, and quantum key distribution \cite{Simon2014012315}. 

\section{Experimental Implementation}
\label{sec:experiment}

\subsection{Xenon-Cavity System}
\label{subsec:xe_cav}

\begin{figure}[t]
\begin{centering}
\centerline{\includegraphics[trim = 20mm 16mm 110mm 35mm, clip, width=0.8\columnwidth]{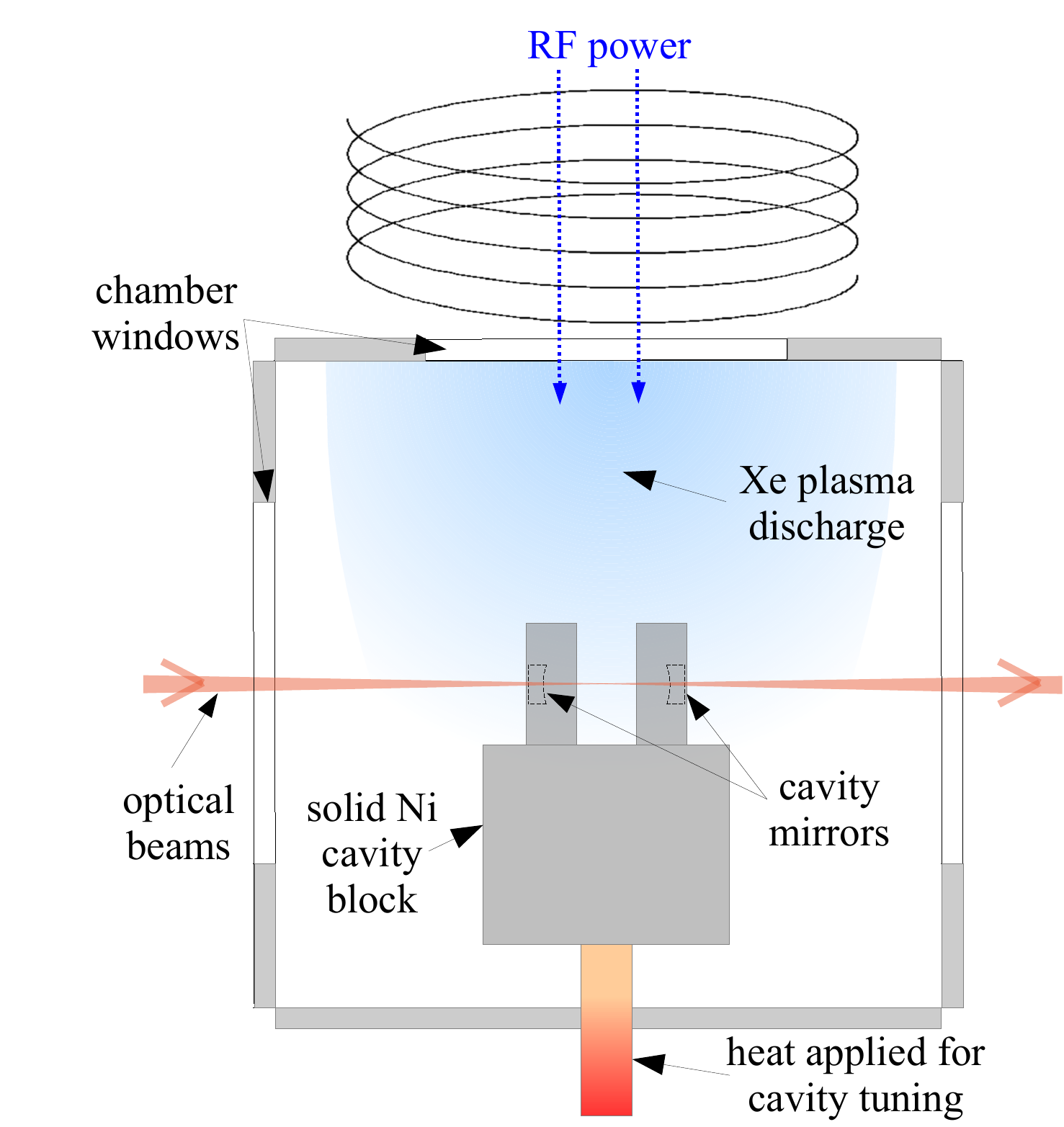}}
\caption{Simplified diagram of the optical cavity filled with metastable xenon atoms. The xenon metastable state is populated through an RF discharge from a resonant circuit (coils and capacitors). Mirrors are held in a solid nickel block and the cavity is frequency tuned by controlling the block's temperature.}\label{fig:Xenon_Cavity}
\end{centering}
\end{figure}

These estimates of the cross-phase modulation attainable with a metastable xenon-filled cavity are promising, but an experimental demonstration is needed to verify them. An apparatus to perform the necessary measurements is currently being constructed. The design of the metastable xenon-filled optical cavity is shown in Fig. \ref{fig:Xenon_Cavity}. Two mirrors are housed in a solid nickel block and placed in a vacuum chamber filled with xenon gas. A resonant RF circuit (capacitors, wire coils) excites the xenon atoms into the metastable state, and the atoms diffuse through the chamber and into the beam path. Tuning of the cavity frequency is accomplished by controlling the temperature of the nickel block. The combined xenon-cavity system creates an optical nonlinearity in which the presence of a single control photon is expected to impart a significant cross-phase modulation on a macroscopic coherent state signal.

\subsection{Experimental Design}
\label{subsec:exp_design}

\begin{figure}[t]
\centerline{\includegraphics[trim = 40mm 175mm 75mm 110mm, clip, width=1.0\columnwidth]{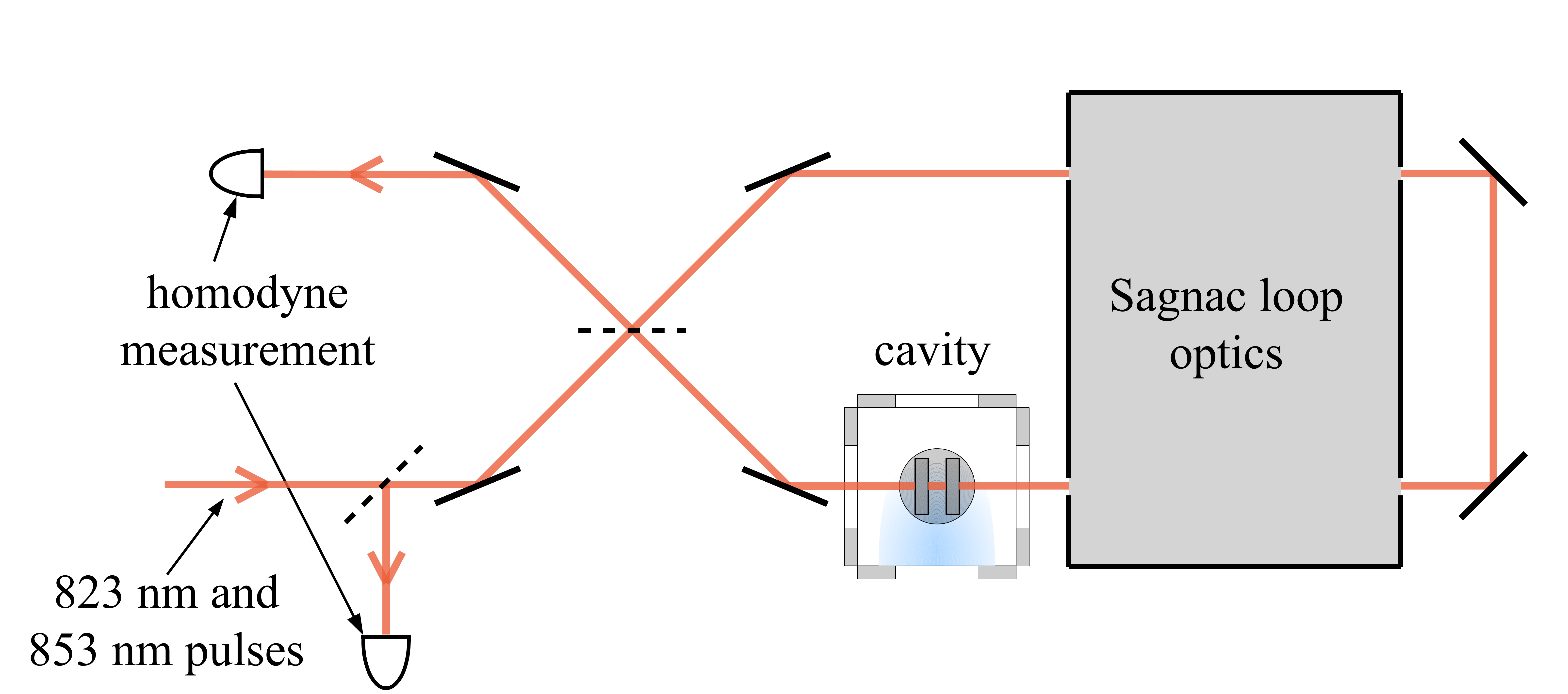}}
\caption{Conceptual layout of experiment to measure single-photon level cross-phase shifts using metastable xenon gas in a high-finesse optical cavity.}\label{fig:Exp_Layout}
\end{figure}

Fig. \ref{fig:Exp_Layout} shows a simplified conceptual diagram of the experimental layout. Control and signal beams at 823 and 853 nm respectively will pass through a pair of electro-optic intensity modulators that are capable of effectively turning the beams on and off at gigahertz rates. Each beam will be divided into separate high intensity and low intensity paths, with the power level controlled by variable in-line fiber optic attenuators. The high intensity beams will saturate the xenon absorption to allow frequency locking to the cavity transmission peak, while the low intensity beams will be required to perform the experiment. Fiber-based optical switches select the high or low intensities.  The laser frequency tuning and switch operation will be controlled in a Labview routine.

During the phase-shift measurements, a light pulse from the signal beam will enter the Sagnac loop containing the optical cavity. A phase modulator will be used to impart a 90$^{\circ}$ phase shift on one but not both of the counterpropagating pulses, maximizing the sensitivity of the interference signal to slight relative phase differences between the two. The clockwise-propagating signal 853 pulse will be attenuated on passing in the reverse direction through an optical isolator, so as not to saturate the xenon $|i\rangle$ to $|h\rangle$ transition, and will serve as the weak signal. The counterclockwise-propagating pulse will not be attenuated and will function as a strong local oscillator in a balanced homodyne measurement at the outputs of the Sagnac loop, which will be used to detect the small cross-phase shift in the weak signal pulse. A single-photon level control pulse at 823 nm will be timed to interact in the cavity with only the clockwise-propagating signal pulse, so that the local oscillator is not affected. Temporal separation of the signal and local oscillator will be accomplished by the addition of an extra 150 m length of optical fiber in the interferometer path.

A beamsplitter will be added into the Sagnac loop to split off and measure the counterclockwise-propagating control pulse. The control amplitude will be reduced such that the average number of photons per pulse is much less than one, and the measurements will then be postselected on the detection of a control photon. The measurements will be immediately repeated with the control turned off, and the results compared to determine the relative phase shift induced by a single control photon.

While the experiment has not yet been completed, we have performed a preliminary demonstration of its feasibility in a measurement of saturated absorption in the metastable $|g\rangle$ to $|i\rangle$ transition. Significant saturation was observed with input powers of only 20 nW \cite{hickman2014}, this corresponds to a strong nonlinearity that is consistent with the expected single-photon cross-phase modulation.

\section{Summary and Conclusions}
\label{sec:summary}

We have proposed a cavity-based approach for the generation and measurement of a weak single-photon cross-phase modulation.
Metastable xenon vapor is used as the nonlinear intracavity medium to avoid the problems associated with the accumulation of rare earth atoms on the optical surfaces \cite{Pittman2013053804}.
The use of bulk xenon also avoids the complexity associated with trapping a single atom, while the use of a high-finesse cavity may allow one to avoid the objections raised in \cite{Shapiro200716, Gea-Banacloche2010043823,Fan2013053601} associated with propagating pulses.

A three level ladder scheme in xenon with transitions at 823 nm and 853 nm is used \cite{Allen1969842,Das2005285}, with detunings added to avoid two photon absorption.
Two different theoretical approaches for describing the cross-phase modulation found that phase shifts of roughly 20 milliradians are possible in this system, and we have given a brief description of experimental progress towards its realization.

Single-photon cross-phase modulation has a number of important applications in quantum optics and quantum information science.
In particular, single-photon cross-phase modulation would allow the generation of macroscopic entangled coherent states which may be useful for quantum-key generation \cite{Simon2014012315} and other applications \cite{Munro2005137,Nemoto2004250502,Jeong2002042305,Kirby2013053822,Kirby2013ARXIV,Enk2001022313,Wang2001022302,Jeong2001052308,Joo2011083601} .
The relative simplicity of the cavity-based approach described here should enable many of these applications in the future.

\section*{Acknowledgements}
The authors would like to acknowledge valuable discussions with C. J. Broadbent and D. E. Jones. This work was supported in part by DARPA DSO under Grant No. W31P4Q-10-1-0018.

%\section*{References}


\begin{thebibliography}{34}
\expandafter\ifx\csname natexlab\endcsname\relax\def\natexlab#1{#1}\fi
\providecommand{\url}[1]{\texttt{#1}}
\providecommand{\href}[2]{#2}
\providecommand{\path}[1]{#1}
\providecommand{\DOIprefix}{doi:}
\providecommand{\ArXivprefix}{arXiv:}
\providecommand{\URLprefix}{URL: }
\providecommand{\Pubmedprefix}{pmid:}
\providecommand{\doi}[1]{\href{http://dx.doi.org/#1}{\path{#1}}}
\providecommand{\Pubmed}[1]{\href{pmid:#1}{\path{#1}}}
\providecommand{\bibinfo}[2]{#2}
\ifx\xfnm\relax \def\xfnm[#1]{\unskip,\space#1}\fi
%Type = Article
\bibitem[{Munro et~al.(2005)Munro, Nemoto, and Spiller}]{Munro2005137}
\bibinfo{author}{W.~J. Munro}, \bibinfo{author}{K.~Nemoto},
  \bibinfo{author}{T.~P. Spiller}, \bibinfo{journal}{New Journal of Physics}
  \bibinfo{volume}{7} (\bibinfo{year}{2005}) \bibinfo{pages}{137}.
%Type = Article
\bibitem[{Nemoto and Munro(2004)}]{Nemoto2004250502}
\bibinfo{author}{K.~Nemoto}, \bibinfo{author}{W.~J. Munro},
  \bibinfo{journal}{Phys. Rev. Lett.} \bibinfo{volume}{93}
  (\bibinfo{year}{2004}) \bibinfo{pages}{250502}.
%Type = Article
\bibitem[{Jeong and Kim(2002)}]{Jeong2002042305}
\bibinfo{author}{H.~Jeong}, \bibinfo{author}{M.~S. Kim},
  \bibinfo{journal}{Phys. Rev. A} \bibinfo{volume}{65} (\bibinfo{year}{2002})
  \bibinfo{pages}{042305}.
%Type = Article
\bibitem[{van Enk and Hirota(2001)}]{Enk2001022313}
\bibinfo{author}{S.~J. van Enk}, \bibinfo{author}{O.~Hirota},
  \bibinfo{journal}{Phys. Rev. A} \bibinfo{volume}{64} (\bibinfo{year}{2001})
  \bibinfo{pages}{022313}.
%Type = Article
\bibitem[{Wang(2001)}]{Wang2001022302}
\bibinfo{author}{X.~Wang}, \bibinfo{journal}{Phys. Rev. A} \bibinfo{volume}{64}
  (\bibinfo{year}{2001}) \bibinfo{pages}{022302}.
%Type = Article
\bibitem[{Jeong et~al.(2001)Jeong, Kim, and Lee}]{Jeong2001052308}
\bibinfo{author}{H.~Jeong}, \bibinfo{author}{M.~S. Kim},
  \bibinfo{author}{J.~Lee}, \bibinfo{journal}{Phys. Rev. A}
  \bibinfo{volume}{64} (\bibinfo{year}{2001}) \bibinfo{pages}{052308}.
%Type = Article
\bibitem[{Joo et~al.(2011)Joo, Munro, and Spiller}]{Joo2011083601}
\bibinfo{author}{J.~Joo}, \bibinfo{author}{W.~J. Munro}, \bibinfo{author}{T.~P.
  Spiller}, \bibinfo{journal}{Phys. Rev. Lett.} \bibinfo{volume}{107}
  (\bibinfo{year}{2011}) \bibinfo{pages}{083601}.
%Type = Article
\bibitem[{Simon et~al.(2014)Simon, Jaeger, and Sergienko}]{Simon2014012315}
\bibinfo{author}{D.~S. Simon}, \bibinfo{author}{G.~Jaeger},
  \bibinfo{author}{A.~V. Sergienko}, \bibinfo{journal}{Phys. Rev. A}
  \bibinfo{volume}{89} (\bibinfo{year}{2014}) \bibinfo{pages}{012315}.
%Type = Article
\bibitem[{Kirby and Franson(2013)}]{Kirby2013053822}
\bibinfo{author}{B.~T. Kirby}, \bibinfo{author}{J.~D. Franson},
  \bibinfo{journal}{Phys. Rev. A} \bibinfo{volume}{87} (\bibinfo{year}{2013})
  \bibinfo{pages}{053822}.
%Type = Article
\bibitem[{Kirby and Franson(2014)}]{Kirby2013ARXIV}
\bibinfo{author}{B.~T. Kirby}, \bibinfo{author}{J.~D. Franson},
  \bibinfo{journal}{Phys. Rev. A} \bibinfo{volume}{89} (\bibinfo{year}{2014})
  \bibinfo{pages}{033861}.
%Type = Article
\bibitem[{Turchette et~al.(1995)Turchette, Hood, Lange, Mabuchi, and
  Kimble}]{Kimble19954710}
\bibinfo{author}{Q.~A. Turchette}, \bibinfo{author}{C.~J. Hood},
  \bibinfo{author}{W.~Lange}, \bibinfo{author}{H.~Mabuchi},
  \bibinfo{author}{H.~J. Kimble}, \bibinfo{journal}{Phys. Rev. Lett.}
  \bibinfo{volume}{75} (\bibinfo{year}{1995}) \bibinfo{pages}{4710--4713}.
%Type = Article
\bibitem[{Venkataraman et~al.(2013)Venkataraman, Saha, and
  Gaeta}]{Gaeta2013138}
\bibinfo{author}{V.~Venkataraman}, \bibinfo{author}{K.~Saha},
  \bibinfo{author}{A.~L. Gaeta}, \bibinfo{journal}{Nat Photon}
  \bibinfo{volume}{7} (\bibinfo{year}{2013}) \bibinfo{pages}{138--141}.
%Type = Article
\bibitem[{Hoi et~al.(2013)Hoi, Kockum, Palomaki, Stace, Fan, Tornberg,
  Sathyamoorthy, Johansson, Delsing, and Wilson}]{Hoi2013053601}
\bibinfo{author}{I.-C. Hoi}, \bibinfo{author}{A.~F. Kockum},
  \bibinfo{author}{T.~Palomaki}, \bibinfo{author}{T.~M. Stace},
  \bibinfo{author}{B.~Fan}, \bibinfo{author}{L.~Tornberg},
  \bibinfo{author}{S.~R. Sathyamoorthy}, \bibinfo{author}{G.~Johansson},
  \bibinfo{author}{P.~Delsing}, \bibinfo{author}{C.~M. Wilson},
  \bibinfo{journal}{Phys. Rev. Lett.} \bibinfo{volume}{111}
  (\bibinfo{year}{2013}) \bibinfo{pages}{053601}.
%Type = Article
\bibitem[{Spani~Molella et~al.(2008)Spani~Molella, Rinkleff, K{\"u}hn, and
  Danzmann}]{Molella2008273}
\bibinfo{author}{L.~Spani~Molella}, \bibinfo{author}{R.-H. Rinkleff},
  \bibinfo{author}{G.~K{\"u}hn}, \bibinfo{author}{K.~Danzmann},
  \bibinfo{journal}{Applied Physics B} \bibinfo{volume}{90}
  (\bibinfo{year}{2008}) \bibinfo{pages}{273--277}.
%Type = Article
\bibitem[{Shiau et~al.(2011)Shiau, Wu, Lin, and Chen}]{Chen2011193006}
\bibinfo{author}{B.-W. Shiau}, \bibinfo{author}{M.-C. Wu},
  \bibinfo{author}{C.-C. Lin}, \bibinfo{author}{Y.-C. Chen},
  \bibinfo{journal}{Phys. Rev. Lett.} \bibinfo{volume}{106}
  (\bibinfo{year}{2011}) \bibinfo{pages}{193006}.
%Type = Article
\bibitem[{Lo et~al.(2011)Lo, Chen, Su, Chen, Chen, Chen, Yu, and
  Chen}]{Chen2011041804}
\bibinfo{author}{H.-Y. Lo}, \bibinfo{author}{Y.-C. Chen},
  \bibinfo{author}{P.-C. Su}, \bibinfo{author}{H.-C. Chen},
  \bibinfo{author}{J.-X. Chen}, \bibinfo{author}{Y.-C. Chen},
  \bibinfo{author}{I.~A. Yu}, \bibinfo{author}{Y.-F. Chen},
  \bibinfo{journal}{Phys. Rev. A} \bibinfo{volume}{83} (\bibinfo{year}{2011})
  \bibinfo{pages}{041804}.
%Type = Article
\bibitem[{Fushman et~al.(2008)Fushman, Englund, Faraon, Stoltz, Petroff, and
  Vu{\v c}kovi{\'c}}]{Fushman2008769}
\bibinfo{author}{I.~Fushman}, \bibinfo{author}{D.~Englund},
  \bibinfo{author}{A.~Faraon}, \bibinfo{author}{N.~Stoltz},
  \bibinfo{author}{P.~Petroff}, \bibinfo{author}{J.~Vu{\v c}kovi{\'c}},
  \bibinfo{journal}{Science} \bibinfo{volume}{320} (\bibinfo{year}{2008})
  \bibinfo{pages}{769--772}.
%Type = Article
\bibitem[{Firstenberg et~al.(2013)Firstenberg, Peyronel, Liang, Gorshkov,
  Lukin, and Vuleti{\'c}}]{firstenberg201371}
\bibinfo{author}{O.~Firstenberg}, \bibinfo{author}{T.~Peyronel},
  \bibinfo{author}{Q.-Y. Liang}, \bibinfo{author}{A.~V. Gorshkov},
  \bibinfo{author}{M.~D. Lukin}, \bibinfo{author}{V.~Vuleti{\'c}},
  \bibinfo{journal}{Nature} \bibinfo{volume}{502} (\bibinfo{year}{2013})
  \bibinfo{pages}{71--75}.
%Type = Article
\bibitem[{Pittman et~al.(2013)Pittman, Jones, and Franson}]{Pittman2013053804}
\bibinfo{author}{T.~B. Pittman}, \bibinfo{author}{D.~E. Jones},
  \bibinfo{author}{J.~D. Franson}, \bibinfo{journal}{Phys. Rev. A}
  \bibinfo{volume}{88} (\bibinfo{year}{2013}) \bibinfo{pages}{053804}.
%Type = Article
\bibitem[{Shapiro and Razavi(2007)}]{Shapiro200716}
\bibinfo{author}{J.~H. Shapiro}, \bibinfo{author}{M.~Razavi},
  \bibinfo{journal}{New Journal of Physics} \bibinfo{volume}{9}
  (\bibinfo{year}{2007}) \bibinfo{pages}{16}.
%Type = Article
\bibitem[{Gea-Banacloche(2010)}]{Gea-Banacloche2010043823}
\bibinfo{author}{J.~Gea-Banacloche}, \bibinfo{journal}{Phys. Rev. A}
  \bibinfo{volume}{81} (\bibinfo{year}{2010}) \bibinfo{pages}{043823}.
%Type = Article
\bibitem[{Fan et~al.(2013)Fan, Kockum, Combes, Johansson, Hoi, Wilson, Delsing,
  Milburn, and Stace}]{Fan2013053601}
\bibinfo{author}{B.~Fan}, \bibinfo{author}{A.~F. Kockum},
  \bibinfo{author}{J.~Combes}, \bibinfo{author}{G.~Johansson},
  \bibinfo{author}{I.-c. Hoi}, \bibinfo{author}{C.~M. Wilson},
  \bibinfo{author}{P.~Delsing}, \bibinfo{author}{G.~J. Milburn},
  \bibinfo{author}{T.~M. Stace}, \bibinfo{journal}{Phys. Rev. Lett.}
  \bibinfo{volume}{110} (\bibinfo{year}{2013}) \bibinfo{pages}{053601}.
%Type = Article
\bibitem[{Walhout et~al.(1994)Walhout, Witte, and Rolston}]{Walhout29942843}
\bibinfo{author}{M.~Walhout}, \bibinfo{author}{A.~Witte},
  \bibinfo{author}{S.~Rolston}, \bibinfo{journal}{Phys. Rev. Lett.}
  \bibinfo{volume}{72} (\bibinfo{year}{1994}) \bibinfo{pages}{2843}.
%Type = Article
\bibitem[{Allen et~al.(1969)Allen, Jones, and Schofield}]{Allen1969842}
\bibinfo{author}{L.~Allen}, \bibinfo{author}{D.~G.~C. Jones},
  \bibinfo{author}{D.~G. Schofield}, \bibinfo{journal}{J. Opt. Soc. Am.}
  \bibinfo{volume}{59} (\bibinfo{year}{1969}) \bibinfo{pages}{842--847}.
%Type = Article
\bibitem[{Das and Karmakar(2005)}]{Das2005285}
\bibinfo{author}{M.~B. Das}, \bibinfo{author}{S.~Karmakar},
  \bibinfo{journal}{The European Physical Journal D - Atomic, Molecular,
  Optical and Plasma Physics} \bibinfo{volume}{32} (\bibinfo{year}{2005})
  \bibinfo{pages}{285--288}.
%Type = Article
\bibitem[{You et~al.(2008)You, Hendrickson, and Franson}]{You2008053803}
\bibinfo{author}{H.~You}, \bibinfo{author}{S.~M. Hendrickson},
  \bibinfo{author}{J.~D. Franson}, \bibinfo{journal}{Phys. Rev. A}
  \bibinfo{volume}{78} (\bibinfo{year}{2008}) \bibinfo{pages}{053803}.
%Type = Book
\bibitem[{Scully(1997)}]{Scully1997}
\bibinfo{author}{M.~O. Scully}, \bibinfo{title}{Quantum optics},
  \bibinfo{publisher}{Cambridge university press}, \bibinfo{year}{1997}.
%Type = Article
\bibitem[{You and Franson(2012)}]{You20121627}
\bibinfo{author}{H.~You}, \bibinfo{author}{J.~Franson},
  \bibinfo{journal}{Quantum Information Processing} \bibinfo{volume}{11}
  (\bibinfo{year}{2012}) \bibinfo{pages}{1627--1651}.
%Type = Article
\bibitem[{Boyd and Gordon(1961)}]{Boyd19611538}
\bibinfo{author}{G.~D. Boyd}, \bibinfo{author}{J.~P. Gordon},
  \bibinfo{journal}{Bell System Technical Journal} \bibinfo{volume}{40}
  (\bibinfo{year}{1961}) \bibinfo{pages}{489--508}.
%Type = Article
\bibitem[{Uhm et~al.(2008)Uhm, Oh, and Choi}]{Uhm2008211501}
\bibinfo{author}{H.~Uhm}, \bibinfo{author}{P.~Y. Oh}, \bibinfo{author}{E.~H.
  Choi}, \bibinfo{journal}{Applied Physics Letters} \bibinfo{volume}{93}
  (\bibinfo{year}{2008}) \bibinfo{pages}{211501--211501--3}.
%Type = Article
\bibitem[{Hilborn(1982)}]{Hilborn1982982}
\bibinfo{author}{R.~C. Hilborn}, \bibinfo{journal}{American Journal of Physics}
  \bibinfo{volume}{50} (\bibinfo{year}{1982}) \bibinfo{pages}{982--986}.
%Type = Article
\bibitem[{Jeong et~al.(2004)Jeong, Kim, Ralph, and Ham}]{Jeong2004061801}
\bibinfo{author}{H.~Jeong}, \bibinfo{author}{M.~S. Kim}, \bibinfo{author}{T.~C.
  Ralph}, \bibinfo{author}{B.~S. Ham}, \bibinfo{journal}{Phys. Rev. A}
  \bibinfo{volume}{70} (\bibinfo{year}{2004}) \bibinfo{pages}{061801}.
%Type = Article
\bibitem[{Gerry(1999)}]{Gerry19994095}
\bibinfo{author}{C.~C. Gerry}, \bibinfo{journal}{Phys. Rev. A}
  \bibinfo{volume}{59} (\bibinfo{year}{1999}) \bibinfo{pages}{4095--4098}.
%Type = Article
\bibitem[{Hickman et~al.(2014)Hickman, Pittman, and Franson}]{hickman2014}
\bibinfo{author}{G.~T. Hickman}, \bibinfo{author}{T.~B. Pittman},
  \bibinfo{author}{J.~D. Franson}  (\bibinfo{year}{2014}).
  \href{http://arxiv.org/abs/1404.0376}{\tt arXiv:1404.0376}.

\end{thebibliography}
\end{document}